\documentclass[osajnl,twocolumn,showpacs,10pt]{revtex4-1} 
\usepackage{amsmath,amssymb,graphicx}

\begin{document}

\title{Beam shaping using genetically optimized two-dimensional photonic crystals}
\author{Denis Gagnon}
\author{Joey Dumont}
\author{Louis J. Dub\'e}\email{Corresponding author: ljd@phy.ulaval.ca}
\affiliation{D\'epartement de physique, de g\'enie physique et d'optique \\Facult\'e des Sciences et de G\'enie, Universit\'e Laval, Qu\'ebec G1V 0A6, Canada}

\begin{abstract}
We propose the use of two-dimensional photonic crystals with engineered defects for the generation of an arbitrary-profile beam from a focused input beam. The cylindrical harmonics expansion of complex-source beams is derived and used to compute the scattered wavefunction of a 2D photonic crystal via the multiple scattering method. The beam shaping problem is then solved using a genetic algorithm. We illustrate our procedure by generating different orders of Hermite-Gauss profiles, while maintaining reasonable losses and tolerance to variations in the input beam and the slab refractive index.
\end{abstract}

\ocis{140.3300, 230.5298, 290.4210} 

\maketitle

\section{Introduction}

Laser beam shaping, defined as redistributing the irradiance and phase of a beam, is of great interest for many applications such as image processing and holography \cite{Dickey2005}, atom guiding \cite{Molina-Terriza2007}, materials processing \cite{Duocastella2012} and controlling random laser emission \cite{Bachelard2012}. Shaping can be achieved using various optical apparatus, such as binary holograms \cite{Brown1966}, conical lenses \cite{Duocastella2012, Herman1991}, solid state lasers \cite{Laabs1996}, and spatial light modulators \cite{Bachelard2012}.
Beam shaping using anisotropic photonic crystals has also been reported \cite{Shadrivov2003, Rasoga2010, Shapira2012}. Moreover, the generation of self-healing, limited-diffraction Bessel-Gauss beams by 2D axicon-shaped photonic crystals has recently been demonstrated by Kurt and Turduev \cite{Kurt2009, Kurt2012}. These promising results highlight the potential of photonic crystal engineering for the generation of beams of \textit{arbitrary} profiles. However, few solutions are available for robust integration of optical elements dedicated to beam shaping on planar lightwave circuits. One of those is the use of a heterogeneous refractive index maps to convert a Gaussian beam to a Bessel-Gauss profile \cite{San-Roman-Alerigi2012}. Nevertheless, planar-waveguide based photonic crystal slabs, consisting of air holes in a high index core, retain immense potential for fabrication of integrated optical elements \cite{Frandsen2004, Pottier2006, Baba2008, DeLaRue2012}.

The aim of this paper is to show that two-dimensional photonic crystals (PhC) can be engineered to achieve \emph{any} specific beam profile required for a given application, while maintaining relatively low scattering losses. Theoretical PhC engineering involves selecting a number of adjustable geometric parameters and performing parametric optimization of a cost function related to the irradiance distribution of the scattered beam. Since the use of more adjustable parameters (usually) results in more diverse output profiles, a fast and accurate numerical method is needed to compute the field scattered by the PhC device. The speed of the method is critical since a large number of configurations must be tested. Consequently, resource-heavy finite-difference time-domain (FDTD) computations \cite{Kurt2012} are not suited for our purpose. We rather use the typically faster multiple scattering computations \cite{Elsherbeni1992, Nojima2005}. The first part of this paper is concerned with a description of the 
scattering 
approach. 
We present a derivation of the cylindrical harmonics expansion of focused beams used to parametrize the wave incident on the PhC. This expansion is required by the multiple scattering formalism.

In the latter part of this paper, we detail the proposed PhC devices and the optimization scheme used. Like Vukovic \textit{et al.} \cite{Vukovic2010}, we choose a basic photonic lattice configuration and allow individual scatterers to be present or absent as the only adjustable parameters, thereby enabling a binary encoding of the configuration space and the use of the standard genetic algorithm (GA) to find the configuration best suited to our purpose \cite{Evans1998, Sivanandam2008}. Our results show that the optimization strategies presented in \cite{Vukovic2010} can be advantageously used to design an integrated beam shaping device. To illustrate this, we present engineered configurations allowing the generation of two different Hermite-Gauss beam profiles with great accuracy, and discuss the  power conversion efficiency of the proposed devices.

\section{Scattering of complex-source beams by PhCs}

This section establishes the theoretical framework used to compute the field scattered by a finite PhC slab. A generic two-dimensional PhC consists of an array of air holes in a planar dielectric waveguide, with a lattice constant of the order of the operation wavelength \cite{Baba2008, Skorobogatiy2009}. Since our goal is to engineer the geometric properties of the PhC to achieve a given beam profile, we only consider finite-size slabs. For modeling purposes, we suppose that every cylinder (hole) is infinite along the axial $z$ direction. The field scattered by the cylinder array is then given by the solution of the 2D Helmholtz equation
\begin{equation}
[\nabla^2 + k^2(x,y)]u(x,y) = 0
\end{equation}
where a harmonic time dependence $\exp(-i\omega t)$ is assumed and $k = k_0 n(x,y)$, where $n$ is the refractive index. Both TM $(u \equiv E_z)$ and TE  $(u \equiv H_z)$ polarized waves can be con\-si\-dered. The wavefunction outside the scatterers can be written as a superposition of an incident and a scattered wave, $u(x,y) = u_i(x,y) + u_s(x,y)$. We then seek the scattered wavefunction $u_s(x,y)$ in the case where $u_i(x,y)$ is a focused beam with a Gaussian shape in the paraxial zone. For this purpose, the incident wavefunction is represented by a complex-source beam (CSB).  This solution has been proposed in order to extend the validity of the Gaussian beam (GB) beyond the paraxial zone \cite{Felsen1975, Heyman2001}. 

Using the Green's function of the inhomogeneous Helmholtz equation for a point source located in the complex plane at coordinates $x'=ix_R$ and $y'=0$, one obtains the CSB solution
\begin{equation}\label{eq:CSB}
u_i(x,y) = H^{(1)}_0 (k r_s)
\end{equation} 
where $H^{(1)}_0$ is a Hankel function of the first kind. The complex distance $r_s$ is given by 
\begin{equation}
r_s \equiv [(y-y')^2 + (x - x')^2]^{1/2}= [y^2 + (x - ix_{R})^2]^{1/2}.
\end{equation}
The complex point-source yields a directional field radiating away from the beam waist ($x=0$). The CSB is continuous everywhere in the real plane except across the branch cut connecting the two singularities at $(x,y) = (0,x_{R})$ and $(x,y) = (0,-x_{R})$. For the purposes of this paper, we shall restrict our attention to scatterers located in the positive $x$ plane, referring the reader to \cite{Mahillo-Isla2008} for regularization strategies in the waist plane. Since the $H_0^{(1)}$ function converges rapidly to a complex exponential, one can readily show that, for $x>0$,
\begin{equation}
u_i(x,y) \sim u_g(x,y)\exp\left(k x_{R} + i\pi/4 \right)
\end{equation}
where 
\begin{equation}
u_g(x,y) = \sqrt{\frac{2}{\pi k (x-ix_{R})}} \exp \left\lbrace ik \left( x + \frac{1}{2} \frac{y^2}{x-ix_{R}} \right)\right\rbrace.
\end{equation}
In other words, the CSB reduces to a GB of Rayleigh distance $x_{R}$ propagating along the $x$ axis in the paraxial zone. Moreover, since the CSB is an analytical solution of the Helmholtz equation exhibiting the cylindrical symmetry characteristic of the multiple scattering method, it is the ideal parametrization of a focused non-paraxial GB incident on an array of cylindrical scatterers.

\subsection{Expansion of complex-source beams in cylindrical harmonics}

To compute the scattered wavefunction via the multiple scattering method, one needs to expand the incident field on a basis of cylindrical waves centered on each individual scatterer. This section is dedicated to the analytic expansion of the aforementioned CSB into cylindrical harmonics. Let $(\rho_n,\theta_n)$ be the cylindrical coordinate system local to the $n^{th}$ scatterer, whose center is located at $(X_n, Y_n)$. We seek a series expansion to rewrite the incident beam in the following fashion
\begin{equation}\label{eq:neumann}
u_i(\rho_n,\theta_n) = \sum_{l=-\infty}^{\infty} a_{nl}^0 J_l(k \rho_n) e^{i l \theta_n}.
\end{equation}
On can rewrite eq. (\ref{eq:CSB}) as 
\begin{equation}
u_i(\rho_n,\theta_n) =  H^{(1)}_0 (k |\boldsymbol{r}_n - \boldsymbol{r}_{sn}|) 
\end{equation}
where $\boldsymbol{r}_n = (\rho_n,\theta_n)$ and $\boldsymbol{r}_{sn}$ is the vector pointing from the center of the $n^{th}$ scatterer to the complex source point at coordinates $(x,y)=(ix_{R},0)$. To uncouple $\boldsymbol{r}_n$ and $\boldsymbol{r}_{sn}$ in the argument of Bessel functions, we apply Graf's addition theorem \cite{Abramowitz1970}. This leads to the following expansion coefficients, similar to those found in \cite{Oguzer1995}
\begin{equation}\label{eq:anl0}
 a_{nl}^0 = (-1)^l H_l(k r_{sn}) e^{-il\mu}
\end{equation}
where 
\begin{equation}\label{eq:rsn}
r_{sn} = \sqrt{(X_n - ix_{R})^2 + Y_n^2}
\end{equation}
and
\begin{equation} \label{eq:mu}
\cos \mu = \frac{X_n - ix_{R}}{r_{sn}}.
\end{equation}
For comparison, the expansion coefficients for a plane wave (PW) incident from the $-x$ axis are given by
\begin{equation}
a_{nl}^0 = i^l e^{ik X_n}.
\end{equation}
The sole knowledge of the $a_{nl}^0$ expansion coefficients of the incident beam allows the use of the multiple scattering method. In a nutshell, one writes the scattered field as a sum of cylindrical waves centered on each individual scatterer
\begin{equation}
u_s(x,y) = \sum_n \sum_l b_{nl} H^{(1)}_l(k \rho_n) e^{i l \theta_n}.
\end{equation}
The matrix equation connecting the expansion coefficients can be written as $s_{nl} a_{nl}^0 = T_{nn'}^{ll'} b_{n'l'}$, with
\begin{equation}
T_{nn'}^{ll'} = \delta_{nn'}\delta_{ll'} - (1 - \delta_{nn'}) e^{i(l'-l) \phi_{n'n}} H^{(1)}_{l-l'}(k R_{nn'})s_{nl}
\end{equation}
where $R_{nn'}$ is the center-to-center distance between scatterers $n$ and $n'$, $\phi_{n'n}$ is the angular position of scatterer $n'$ in the frame of reference of scatterer $n$ and $s_{nl}$ is a constant resulting from the application of electromagnetic boundary conditions. Further details are given in \cite{Elsherbeni1992, Nojima2005}.

Remarkably, except for the computation of a cylindrical function, no supplementary numerical cost is involved in computing the scattered wavefunction in the case of an incident CSB rather than an incident PW. Indeed, the core operation of the multiple scattering method involves computing the $b_{nl}$ coefficients via a matrix inversion, whose computation scales as the square of the number of scatterers $N_s$, regardless of the shape of the incident beam. However, a simple analysis shows that the convergence of (\ref{eq:anl0}) is limited to a disk not intersecting or touching the branch cut between $(x,y)=(0,x_{R})$ and $(x,y)=(0,-x_{R})$. In other words, scatterers must not intersect or touch the branch cut for the expansion to be used in scattering computations. This restriction is not present in the case of an incident PW. It does not restrict the scope of 
our computations since we position all scatterers in the $+x$ half plane.

\section{Beam shaping computations}

\begin{figure}
\centering
\includegraphics{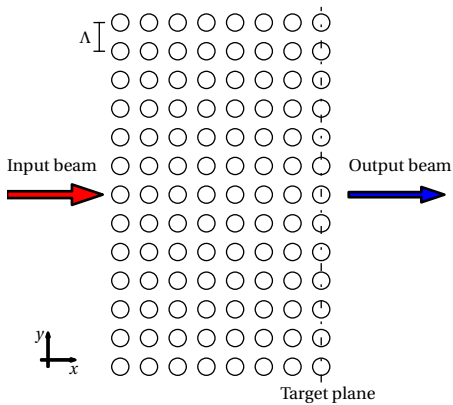}
\caption{(Color online) Basic photonic lattice configuration ($N_s=104$). To generate a desired beam profile, defects can be present or absent. We impose a vertical mirror symmetry, resulting in $2^{56}$ possible configurations. The dotted line indicates the plane used for the computation of the desired beam profile.}\label{fig:geometry}
\end{figure}

\begin{figure}
\centering
\includegraphics{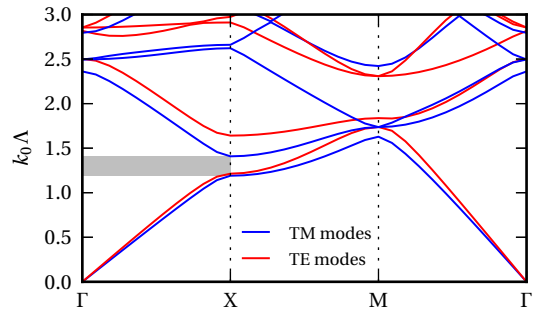}
\caption{(Color online) Band structure for a square lattice of air holes of diameter $D=0.6\Lambda$ in a dielectric medium with refractive index 2.76. The location of the partial bandgap is shaded. Eigenmodes were computed using the \textsc{MIT Photonic bands} software package \cite{Johnson2001}. }\label{fig:bandgap}
\end{figure}

\subsection{Problem definition}
The objective is to find a PhC configuration which, when illuminated with a CSB, produces a scattered wavefunction that matches a desired irradiance profile in a given plane. Let $\bar{u}(x,y)$ be the desired output wavefunction. The beam shaping problem can be formulated as the minimization of the following integral
\begin{equation}
I(x_0) = \dfrac{\int \big||u(x_0,y)|^2 - |\bar{u}(x_0,y)|^2 \big| dy}{\int |\bar{u}(x_0,y)|^2 dy } 
\end{equation}
where $x_0$ is the location of the target plane. This is equivalent to minimizing the root sum of squares (RSS)  of irradiance variations at a set of points of the target plane \cite{Dickey2005}. It is worth noting that we do not take into account the phase of the output beam, only the amplitude. This increases the number of ``acceptable'' configurations in the problem space, at the cost of losing information about the collimation of the output beam in the optimization process. Large variations in the output phase front may result in large output beam divergences, although this is not critical for applications such as materials processing \cite{Dickey2005, Duocastella2012}. Moreover, since backscattering losses are mostly unavoidable in PhC devices, imposing a peak irradiance value is too severe a condition for the optimization algorithm. We rather seek a normalized irradiance profile, and evaluate backscattering losses \textit{a posteriori}.   

The basic scatterer geometry (fig. \ref{fig:geometry}) is a variation of that presented in \cite{Xing2005, Vukovic2010}, i.e. part of a square lattice of air holes embedded in a medium of index $n = 2.76$. The diameter of all holes is set to $D = 0.6 \Lambda$, where $\Lambda$ is the lattice constant. The infinite counterpart of this photonic lattice exhibits a partial photonic bandgap for both polarizations in the $\Gamma - \mathrm{X}$ direction (see fig. \ref{fig:bandgap}). Although the strong confinement associated with a full photonic bandgap is exploited in the case of waveguide design \cite{Xing2005}, it is not mandatory for beam shaping purposes. Indeed, the purpose of the finite PhC slab is not to act as a Bragg reflector, but rather to redistribute the incident beam irradiance via multiple scattering. We shall therefore concentrate on operating wavelengths near the partial bandgap to ensure relatively strong scattering.

For definiteness, we prescribe our incident beam as a TM-polarized CSB given by (\ref{eq:CSB}) with a half-width $w_0=2.5\Lambda$ and a wavenumber $k_0 = 1.76 / \Lambda$ for a Rayleigh distance $x_R =k_0 w_0^2 /2 = 5.48 \Lambda$.  Although the desired output beam and target plane can be arbitrary, for illustrative purposes we have chosen to generate Hermite-Gauss beam profiles of half-width $w$ at the device output, that is
\begin{equation}
|\bar{u}_m(x_0,y)|^2 = \left[\mathcal{H}_m\left( \xi \right) \right]^2 \exp \left( -\xi^2  \right)
\end{equation}
where $\xi = \sqrt{2} y / w$ and $\mathcal{H}_m(\xi)$ is a Hermite polynomial. The first two orders are
\begin{equation}
\begin{aligned}
\mathcal{H}_1(\xi) &= 2\xi \\
\mathcal{H}_2(\xi) &= 4\xi^2 - 2
\end{aligned}
\end{equation}
while $\mathcal{H}_0(\xi) = 1$. For simplicity, we require further that the half-width $w$ of the desired beam profile be identical to $w_0$. 

\begin{figure}
\centering
\includegraphics{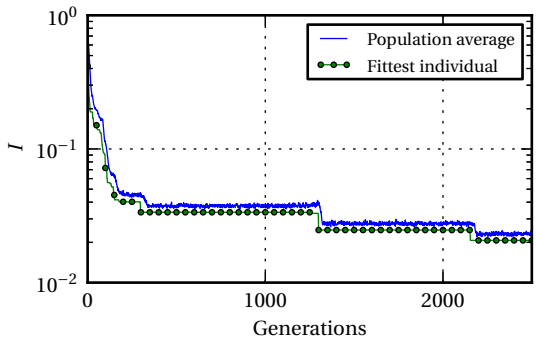}
\caption{(Color online) Convergence of the standard GA used to find the configuration shown on fig \ref{fig:hermite1}. The fitness value reached is $1/I \sim 47.6$.}\label{fig:evolution}
\end{figure}

We use a genetic algorithm (GA) to find the configuration best suited to the generation of a given beam profile \cite{Evans1998, Vukovic2010}. The problem encoding is binary, with each configuration being assigned a ``genotype'' of length equal to the number of available scatterer sites. For the purpose of demonstration, we have targeted symmetric beam shapes and have explicitly imposed mirror symmetry of the scatterers about the $y$-axis. This effectively reduces the problem space dimension, but the method is equally efficient for asymmetric beam shapes. Each trial configuration is assigned a fitness value inversely proportional to $I$. Populations of 200 individuals are generated and evolution takes place until an optimum is reached, typically within a few thousand generations (see fig. \ref{fig:evolution}). We use the standard GA evolutionary operators: roulette wheel sampling, mutation probability $p_m = 0.002$, 
uniform crossover with probability $p_c = 0.2$ and elitism. It is noteworthy that the computation of the fitness function, which implies a matrix inversion and field evaluation via the multiple scattering method, takes only a few seconds for one generation (200 configurations).

\subsection{Generation of beam profiles and tolerance of configurations}

\begin{figure}
\centering
\includegraphics{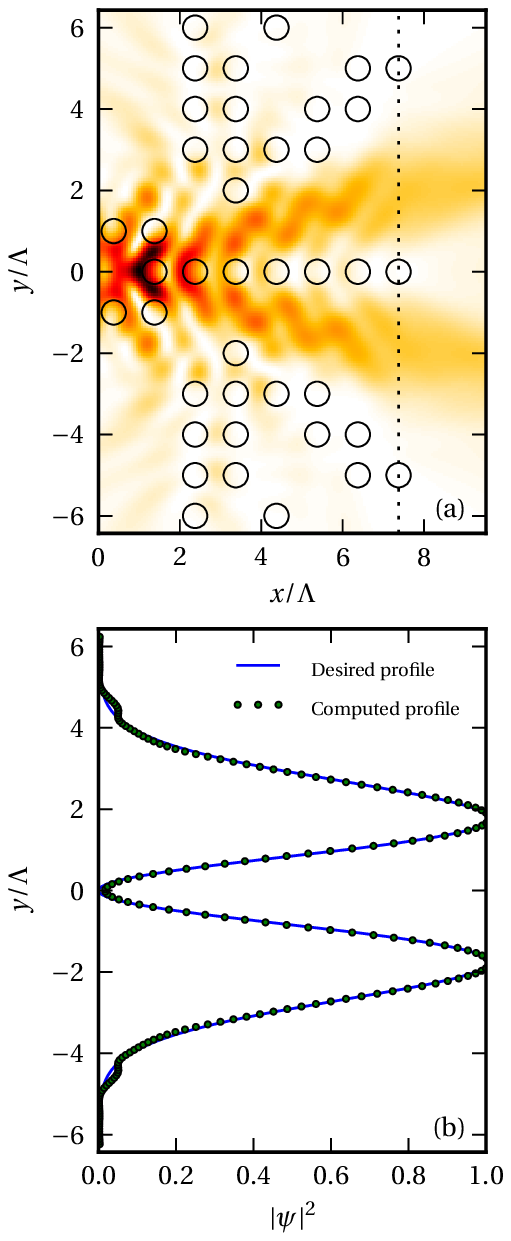} 
\caption{(Color online) Generation of order 1 Hermite-Gauss beam. (a) Optimized configuration and field profile ($N_s=41$). The target plane is indicated by a dashed line. (b) Comparison of computed irradiance along target plane and desired profile (arbitrary units). This design is characterized by $I=0.021, \eta = 0.705$.}\label{fig:hermite1}
\end{figure}

\begin{figure}
\centering
\includegraphics{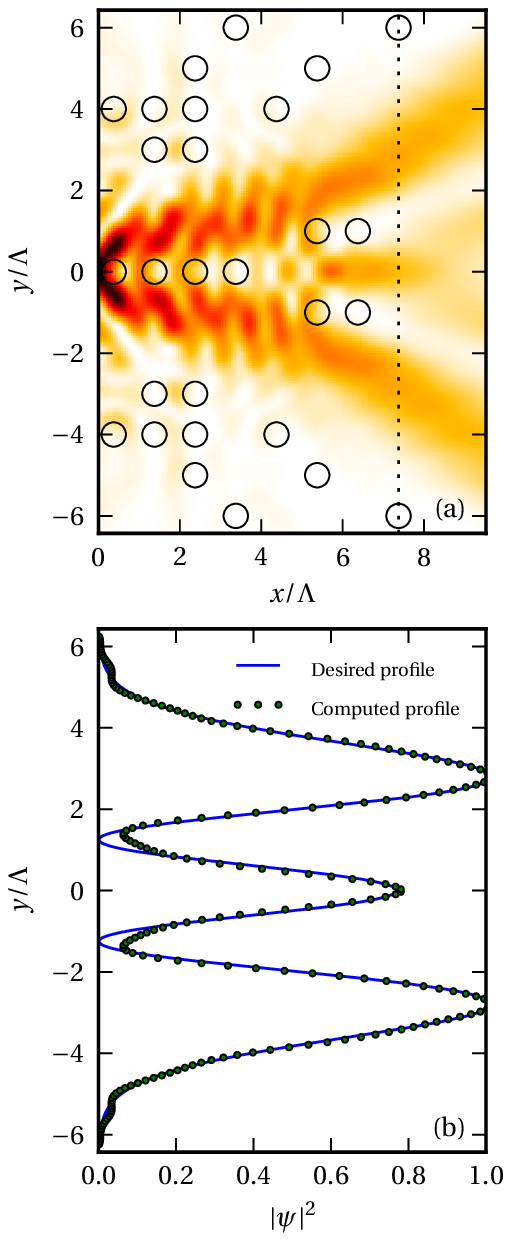}
\caption{(Color online) Generation of order 2 Hermite-Gauss beam. (a) Optimized configuration and field profile ($N_s=28$). The target plane is indicated by a dashed line. (b) Comparison of computed irradiance along target plane and desired profile (arbitrary units). This design is characterized by $I=0.044, \eta = 0.785$.}\label{fig:hermite2}
\end{figure}

\begin{figure}
\centering
\includegraphics{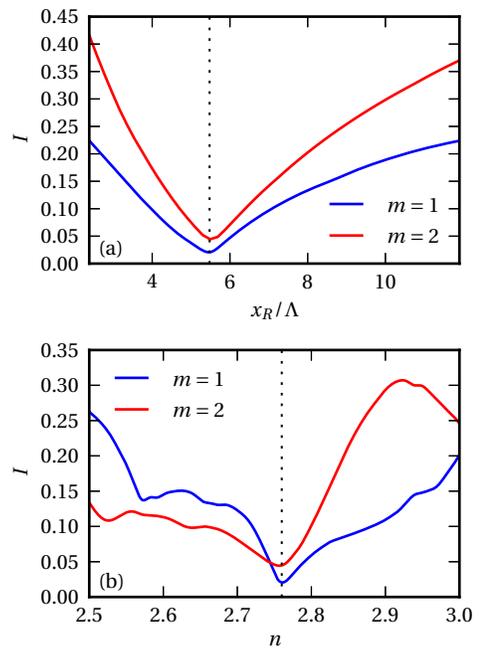}
\caption{(Color online) Tolerance of PhC lattice configurations to (a) variations of the Rayleigh distance of the input beam and (b) group refractive index of the slab. The design values are indicated by a dotted line.}\label{fig:sweep}
\end{figure}

In this section, we present the best configurations found for order 1 and 2 Hermite-Gauss beam profiles, exhibiting a zero and a maximum on the propagation axis, respectively. Results shown on figs. \ref{fig:hermite1} and \ref{fig:hermite2} highlight the possibility to generate order 1 and 2 Hermite-Gauss beam profiles with great accuracy ($I < 0.05$) and are representative of a number of calculations that we have performed. For comparison, the error on the amplitude profile for the PhC device reported in \cite{Rasoga2010} is around 10 \%, while the error of the integrated device proposed in \cite{San-Roman-Alerigi2012} is around 5 \%. This shows that our designs perform equally well or better than recently proposed integrated beam shaping solutions with respect to the profile accuracy. We also stress that the method used is not limited to a single lattice nor to a specific output beam profile. For example, we have obtained profiles with similar accuracy using a triangular lattice with the same refractive 
index. 

Since our primary goal is to obtain an accurate normalized profile via GA optimization, the best configurations found do not necessarily exhibit low backscattering losses. To quantify these losses, we compute the efficiency $\eta$ of the best designs by evaluating the ratio between the electromagnetic power transmitted in the target plane and the total incident power; that is
\begin{equation}
\eta = \dfrac{\int_{-\infty}^{\infty} S_x(x_0,y) dy }{\int_{-\infty}^{\infty} S_x(x_{in},y) dy }
\end{equation}
where $x_{in}$ is the location of the input plane and $S_x$ is the $x$ component of the time-averaged Poynting vector \cite{Nojima2005}. The computation of $\eta$ is achieved via numerical quadrature. As our computations show, efficiencies of optimized configurations typically fall between 70 \% and 80 \%. These numbers are only 10-20 \% smaller than proposed integrated beam shaping devices \emph{specifically} tailored for high efficiencies: references \cite{Rasoga2010, San-Roman-Alerigi2012} report efficiencies of $\sim 90$ \%. It is therefore quite rewarding that our final configurations not only provide a high profile accuracy, but also a low loss design. Of course, if a higher efficiency is critical to a given application, it is always possible to alter the fitness function of the GA to optimize for efficiency as well.

It is instructive to examine the tolerance of optimized PhC configurations to variations of the design parameters. In experimental situations, the Rayleigh distance may vary if the input beam focusing is more or less controlled. On the other hand, the slab refractive index may be fixed using the effective index approximation \cite{Qiu2002}. To assess the tolerance to variations of these two parameters, we have computed the RSS integral $I$ for various values of $x_{R}/ \Lambda$ and $n$ around the design values, while maintaining all others parameters fixed (fig. \ref{fig:sweep}). Results show that varying the value of $x_{R}/\Lambda$ by $\pm 1$, one full lattice spacing, preserves the low value of $I$ (under 0.10), especially in the case of the order 1 Hermite-Gauss beam profile. The PhC configurations presented are also robust with respect to the parameter $n$. It is possible to draw two observations from these computations. First, it is not necessary to run a GA search over a wide range of parameters to 
keep the fitness of the PhC designs within acceptable limits of performance even if some parameters are only approximately known in experimental applications. Second, the results show that the fabrication of a PhC based integrated beam shaper operating in the infrared ($\lambda_0 \sim 1500$ nm, $\Lambda \sim$ 500 nm) is well within reach of current fabrication techniques. Indeed, devices operating in that regime have been successfully fabricated in silicon-on-insulator material using UV lithography \cite{Frandsen2004, Pottier2006, Baba2008, DeLaRue2012}.

\section{Conclusion}

In this paper, we have presented a general design method based on a genetic algorithm for beam shaping using integrated two-dimensional photonic crystals. Parametrization of the incident Gaussian-like beam was achieved using the CSB solution of the Helmholtz equation. The cylindrical harmonics expansion of the incident CSB allows for the use of the multiple scattering method to compute the field scattered by the PhC slab. This method enables fast computation of the amplitude profile of the beam scattered by individual photonic lattice configurations. 

Using this design method, we have tailored photonic crystal devices for the conversion of a CSB to order 1 and 2 Hermite-Gauss beam profiles. The associated beam shaping error ($< 5 \%$) compares advantageously to other known integrated solutions. We also found that over 70 \% of the input beam power was channeled to the output beam. Although we have used a square lattice and required a Hermite-Gauss profile, different lattices and output beam profiles can be accommodated at will.

We have also evaluated the sensitivity of the output beam to variations in the depth of focus of the input beam and the slab refractive index. Our results show that integrated amplitude beam shapers may very well be fabricated using current technology.

\section*{Acknowledgments}
The authors acknowledge financial support from the Natural Sciences and Engineering Research Council of Canada (NSERC). DG is supported by a NSERC Postgraduate Scholarship. The authors also thank R. Dub\'e-Demers and G. Painchaud-April for useful discussions.

\bibliographystyle{osajnl}

\end{document}